\documentstyle[aps,epsf,12pt]{revtex}

\newcommand{\be}{\begin{equation}}
\newcommand{\ee}{\end{equation}}
\newcommand{\bea}{\begin{eqnarray}}
\newcommand{\eea}{\end{eqnarray}}

\tightenlines 

\begin{document}
\draft
\title{Note on flat foliations of spherically symmetric spacetimes}
\author{Viqar Husain$^*$, Asghar Qadir$^\#$ and Azad A. Siddiqui$^\dagger $}
\address{\baselineskip=1.4em $^*$Department of Mathematics and Statistics,\\
University of New Brunswick, Fredericton, NB Canada E3B 5A3. \\
$^\#$Department of Mathematical Sciences,\\
King Fahd University of
Petroleum and Minerals, Dhahran 31261, Saudi Arabia. \\
$^\dagger$E\&ME College,\\ National University of Science and Technology, 
Rawalpindi, Pakistan.}
\maketitle
\bigskip 

\begin{abstract}
\noindent It is known that spherically symmetric spacetimes admit flat 
spacelike foliations. We point out a simple method of seeing this result 
via the Hamiltonian constraints of general relativity. The method yields 
explicit formulas for the extrinsic curvatures of the slicings.
\end{abstract}

\preprint{\small}

\bigskip

The Painleve-Gulstrand form \cite{P,G} of the metric of Schwarzschild
spacetime 
\begin{equation}
ds^2 = -\left( 1 - {\frac{2M }{r}}\right) dt^2 + 2\sqrt{{\frac{2M}{r}}}\ dtdr
+ dr^2 + r^2 d\Omega^2.
\end{equation}
demonstrates that this spacetime has a foliation by flat spatial surfaces.
This generalizes to all spherically symmetric and time dependent 
spacetimes, for which the metric may be written as
\be 
ds^2 = -f^2(r,t)\ dt^2 + 2g(r,t)\ dtdr + dr^2 + r^2 d\Omega^2.
\ee

The question of what observers in the spacetime see flat spacelike constant
time surfaces was addressed recently in \cite{aq}: the surfaces of the
foliation are orthogonal to the trajectories of observers freely falling
from rest at infinity. (Such flat foliations have also been used to study
Hawking radiation \cite{kw,cj}.) A similar result holds for the
Reissner-Nordstrom spacetime \cite{aq}.

The purpose of this note is to demonstrate a method of showing that all
spherically symmetric spacetimes admit flat spacelike foliations via the
Hamiltonian equations of general relativity. The approach involves finding
solutions, in spherical symmetry, of the Hamiltonian and spatial
diffeomorphism constraints of general relativity 
\begin{eqnarray}
{\cal H} &\equiv& {\frac{1}{\sqrt{q}}} G_{abcd}\tilde{\pi}^{ab}\tilde{\pi}%
^{cd} - \sqrt{q}^{(3)} R + \tilde{\rho} = 0,  \label{Ham} \\
{\cal C}^a &\equiv& \partial_b\tilde{\pi}^{ba} + \tilde{P^a} = 0,
\end{eqnarray}
where $(q_{ab},\tilde{\pi}^{ab})$ are the canonically conjugate Hamiltonian
variables, $\tilde{\rho}$ and $\tilde{P_a}$ are the matter energy density and
momentum, $G_{abcd} = (g_{ac}g_{bd} + g_{ad}g_{bc} - g_{ab}g_{cd})/2$ is the
DeWitt supermetric, $D_a$ is the covariant derivative of $q_{ab}$, and $%
\tilde{\ }$ denotes densities of weight one.

The constraints are first class, therefore they continue to be satisfied on
all time slices under evolution. Furthermore the lapse and shift can be
chosen such that flat slices evolve to flat slices. This is easily done 
by solving the Hamiltonian equations for the lapse and shift such that the 
time derivative of the spatial metric is zero, ie. that it remains flat. 
Thus it is sufficient to find flat slice initial data in spherical symmetry 
to show that there exist flat slice foliations.\footnote{After this work was 
submitted for publication, we learned of Ref. \cite{gm} where an initial value 
approach for flat slice foliations is also studied. Our approach stems from 
the slightly earlier work \cite{vh}.}  

Following \cite{vh}, we assume that a flat slice foliation exists, and seek
solutions of the constraints. Let $q_{ab} = e_{ab}$, the flat Euclidean
3-metric in (global) rectangular coordinates $x^a$. The general form of the
conjugate momentum is given by 
\be
\tilde{\pi}^{ab}=f(r)\ n^a n^b+g(r)\ e^{ab}.
\ee
where $f(r)$ and $g(r)$ are functions to be determined, 
$n^{a}=x^{a}/r$ is the unit radial vector and $r^{2}=x^{a}x^{b}e_{ab}$. 
That this is the most general form for $\tilde{\pi}^{ab}$ is easy to see 
by noting that the symmetric $\tilde{\pi}^{ab}$ must be constructed from 
the only objects at hand given flat slicing and spherical 
symmetry: $n^a$ and $e_{ab}$.

Substituting these into the constraints gives 
\bea 
(f - 3g)(f+g) + \tilde{\rho} &=& 0, \\
f^{\prime}+ g^{\prime}+ {\frac{2f}{r}} + \tilde{P}^r &=& 0,
\label{gen0}
\eea
where $\tilde{P}^r$ is the radial matter momentum. These are the two 
equations of interest: the unknown functions are ($f(r),g(r)$), with the 
given matter variables $(\tilde{\rho},\tilde{P^r})$. These equations 
may be rewritten in a form more emanable for general solution by replacing $g(r)$ 
with the canonical variable $\tilde{\pi}(r) := e_{ab}\tilde{\pi}^{ab}=f+3g$,  
and solving the algebraic constraint: 
  
\begin{eqnarray}
f &=& \pm\ {\frac{1}{2}}\ \sqrt{\tilde{\pi}^2 - {3\tilde{\rho}}},  \label{H}
\\
f^{\prime}&+& {\frac{3f}{r}} + {\frac{\tilde{\pi}^{\prime}}{2}} + {\frac{3}{2%
}} \tilde{P}^r = 0.  \label{D}
\end{eqnarray}
Finally, the substitution $\tilde{\pi}(r)=\sqrt{3\tilde{\rho}(r)}\ {\rm cosh}z(r)$
gives $f(r) = \pm  \ \sqrt{3\tilde{\rho}}\ {\rm sinh}z(r)/2$, and  
Eqn. (\ref{D}) becomes 
\be 
 \pm z^{\prime }+{\frac{3}{2r}}\ \left( 1-e^{\mp 2z}\right) 
+ e^{\mp z}\ {\frac{\sqrt{3}%
\tilde{P}^{r}}{\sqrt{\tilde{\rho}}}}+{\frac{\tilde{\rho}^{\prime }}{2\tilde{%
\rho}}} = 0.  \label{gen}
\ee
Thus the problem is reduced to solving Eqn. (\ref{gen}). This nonlinear ode can 
be solved at least numerically for given matter energy-momentum. In particular, 
for $\tilde{P}^{r}=0$ this gives a rather explicit relationship between 
$\tilde{\rho}$ and $z$, and hence $\tilde{\pi}$.

It is worth mentioning that a problem can arise since $\tilde{\rho}$ is a
function of $r$, so that the expression inside the square-root in (\ref{H})
can become negative. This occurs, for example, in the foliation of the
Reissner-Nordstrom geometry. It was dealt with in \cite{aq} by computing the
hypersurface in the domain where the expression remains non-negative and
then continuing numerically across the boundary (where it becomes negative)
by using a Taylor expansion. This procedure may be adopted for other spacetimes 
where this problem occurs, including cases where there is more than one such 
boundary. In the following we discuss some special cases of interest, where the 
solutions are {\it explicit.}  

\medskip

\noindent (i) \underbar{{\it Schwarzschild spacetime:}} $(\tilde{\rho} = 
\tilde{P^r}=0).$ This case is most easily solved using (\ref{gen0}). 
The solution is $f = 3g$ with $f = Cr^{-3/2}$ where $C$ is
an integration constant. 
This constant is related to the Schwarzschild mass $M$ by $C\sim \sqrt{M}$. 
Note that $f=-g$ also solves the Hamiltonian
constraint, but this gives zero 
extrinsic curvature which is the Minkowski
solution.

\medskip

\noindent (ii) \underbar{{\it Reissner-Nordstrom spacetime:}} ($\tilde{\rho}
=Q^{2}/r^{4}$, and $\tilde{P^{r}}=\tilde{E}^{r}\partial _{r}A_{r}=0$ since
the spatial part of the vector potential $A_{a}$ vanishes for this metric). 
Eqn. (\ref{gen}) reduces to 
\begin{equation}
z^{\prime }\mp {\frac{1}{2r}}\left( 3e^{\mp \ 2z}+1\right) =0.
\end{equation}
This has general solution 
\begin{equation}
z(r)=\pm \ {\frac{1}{2}}\ln (Cr-3).
\end{equation}
Thus 
\begin{equation}
\tilde{\pi}={\frac{\sqrt{3}Q}{r^{2}}}\cosh z={\frac{\sqrt{3}Q}{2r^{2}}}\ {%
\frac{Cr-2}{\sqrt{Cr-3}}}.
\end{equation}
With the integration constant $C$ written as $C=6M/Q$, the solutions are 
\begin{equation}
f_{\pm }=\pm \frac{\sqrt{\tilde{\pi}^{2}-3Q^{2}/r^{4}}}{2},\ \ \ \ \ g_{\pm
}=\frac{\tilde{\pi}}{3}\pm \frac{\sqrt{\tilde{\pi}^{2}-3Q^{2}/r^{4}}}{2},
\end{equation}
This result agrees with the forms given in 
\cite{aq} obtained by constructing the flat slice foliation of the
Reissner-Nordstrom metric via the use of freely falling frames. \medskip 

\noindent(iii) \underbar{{\it deSitter spacetime:}} ($\tilde{\rho}=\sqrt{q}
\Lambda >0$, $\tilde{P}^r = 0$). Eqn. (\ref{gen}) is  
\be 
z^{\prime}\mp {\frac{3}{2r}} \left( e^{\mp\ 2z} - 1 \right) = 0,
\ee 
which has general solution 
\begin{equation}
z = \pm{\rm ln}\left( \sqrt{1-K/r^3} \right), \ \ \ \ \ (K> 0).
\end{equation}
Thus 
\begin{equation}
\tilde{\pi} = \pm{\frac{1}{2}}\sqrt{{\frac{3\Lambda r^3}{r^3-K}}}.
\end{equation}

\medskip \noindent(iv) \underbar{{\it Anti-deSitter spacetime:}} ($\tilde{
\rho}=-\sqrt{q}/l^2$, $\tilde{P}^r = 0$). Eqn. (\ref{gen}) may be rewritten as
\begin{equation}
(1\pm {\rm sin}z)\ z^{\prime}\pm 3\ {\frac{{\rm cos}z}{r}} = 0.
\end{equation}
This has solution 
\begin{equation}
z = {\rm arctan} \mp(1-Kr^3), \ \ \ \ (K>0),
\end{equation}
which gives 
\begin{equation}
\tilde{\pi} = \mp{\frac{\sqrt{3}}{l}} (1-Kr^3).
\end{equation}

In summary, we have given a simple method for explicitly finding flat slice
initial data for arbitrary spherically symmetric spacetimes, with the
general case summarized in Eqn. (\ref{gen}). The method also gives a way to
verify that all spherically symmetric spacetimes have flat slice foliations,
via existence of solutions of Eqn. (\ref{gen}). 

\bigskip

\noindent {\it Acknowledgements:} We are grateful to the 25th. International 
Nathiagali Summer College on Physics and Contemporary Needs, where this work 
was initiated. Work of VH was supported in part by the Natural
Science and Engineering Research Council of Canada. AQ is
grateful to KFUPM for facilities provided there.

\end{document}